\documentclass[aps,prd,pdflatex,reprint,10pt,secnumarabic,amssymb,nobibnotes]{revtex4-2}
\usepackage{amsmath}
\usepackage{amssymb}
\usepackage{amsthm}
\usepackage{bbm}
\usepackage{graphicx}
\usepackage{hyperref}
\usepackage{physics}
\usepackage{systeme}
\usepackage{times}
\usepackage{epstopdf}
\usepackage{float}
\usepackage{xcolor}
\usepackage[normalem]{ulem}

\hypersetup{
	colorlinks=true,linkcolor=blue,citecolor=blue,
	filecolor=blue,urlcolor=blue,breaklinks=true
}

\begin{document}
	
\title{Quantum heat machines enabled by twisted geometry}

\author{Cleverson Filgueiras}
\email{cleverson.filgueiras@ufla.br}
\affiliation{Departamento de F\'{i}sica, Universidade Federal de Lavras\\
	Caixa Postal 3037, 37200-000 Lavras-MG, Brazil
      }
  
\author{Moises Rojas}
\email{moises.leyva@ufla.br}
\affiliation{Departamento de F\'{i}sica, Universidade Federal de Lavras\\
	Caixa Postal 3037, 37200-000 Lavras-MG, Brazil
      }
 
\author{Edilberto O. Silva}
\email{edilberto.silva@ufma.br}
\affiliation{
        Departamento de F\'{i}sica,
        Universidade Federal do Maranh\~{a}o,\\
        65085-580, S\~{a}o Lu\'{i}s, Maranh\~{a}o, Brazil
      }
  \author{Carlos Romero}
  \email{cromero@fisica.ufpb.br}
  \affiliation{
  Departamento de F\'{i}sica, Universidade Federal da Para\'{i}ba\\ Caixa Postal 5008, 58059-970, João Pessoa-PB, Brazil
  }
\date{\today }
\begin{abstract}
In this paper, we analyze the operation of an Otto cycle heat machine driven by
a non-interacting two-dimensional electron gas on a twisted geometry. We show that due to both the energy quantization on this structure and the adiabatic transformation of the number of complete twists per unit length of a helicoid, the machine performance in terms of output work, efficiency, and operation mode can be altered. We consider the deformations as in a spring, which is either compressed or stretched from its resting position. The realization of classically inconceivable Otto machines, with an incompressible sample, can be realized as well. The energy-level spacing of the system is the quantity that is being either compressed or stretched. These features are due to the existence of an effective geometry-induced quantum potential which is of pure quantum-mechanical origin.
\end{abstract}
\keywords{Geometry, 2DEG, Quantum Thermodynamics, Heat Machines, Otto Cycle}

\maketitle

\section{\label{sec:level1}Introduction}
A quantum heat engine (QHE) is a device that converts heat energy into work, taking a quantum system as its work substance \cite{first}. It consists of a sequence of transformations(strokes)in Hilbert´s space \cite{PhysRevE.76.031105,PhysRevE.79.041129}. This topic has been an active research branch over the last few years.  Many conceptual designs of the such engine were proposed. For example, magnetically driven QHE's efficiency was addressed in \cite{PhysRevE.89.052107}.  A better performance for a graphene-based QHE, driven by a superposition of magnetic field and a mechanical strain, was pointed out in \cite{PhysRevE.91.052152}.  The performance of a QHE due to the degeneracy in the Landau levels was investigated in \cite{e19120639}.  The investigation of the efficiency of QHE's in power-law trapping potentials can be viewed in \cite{PhysRevE.90.012145}. Also, quantum entanglement has potential applications in quantum communications and in information processing \cite{PhysRevLett.87.017901,PhysRevA.64.012313}. Therefore, it is quite interesting to investigate the quantum heat engines working with quantum entangled systems. In this context, different types of interaction between the subsystems of the working substance have been employed, among them, the Heisenberg XXX model \cite{PhysRevA.75.062102}, the model with XX interaction
\cite{PhysRevE.83.031135} and Dzyaloshinski-Moriya interaction \cite{Zhang2008}.  Obviously, many other examples can be found in the literature since we have a wide range of quantum systems that can be evoked as working substances of QHE's \cite{PhysRevE.94.022109,critical,PhysRevE.104.014149}. The QHE is a subject inside the field of {\it quantum thermodynamics}. For an introduction to this topic, we refer to Refs. \cite{e18050173,doi:10.1080/00107514.2016.1201896}.

Another active branch of research over the last years is the investigation of the dynamics of non-relativistic quantum particles constrained to move on curved surfaces. The Schr\"{o}dinger equation for this case was derived in \cite{PRA1981}. The key ingredient revealed by it is the geometry-induced quantum potential, which is of a quantum nature solely. Among the vast applications which followed such a work, we can find the fundamental ones that set the basis of the theory based on the thin-layer quantization approach. The inclusion of external electric and magnetic fields was implemented in \cite{ferrari}. In it, no coupling between the fields and the quantum geometric potential arises. The version of the Pauli equation for a charged spin particle on a curved surface in external electric and magnetic fields was addressed in \cite{paulicurved}. The refinement of the fundamental framework of the thin-layer quantization procedure considering the surface thickness can be found in \cite{Wang201668}. The effects due to electronic effective masses in curved structures were pointed out in \cite{souza2018curved,position}. By examining these references, many applications of such an approach can be found. The ability to build two-dimensional curved substrates in various shapes is a reality \cite{shapes,experiment}.
As we have said, the applications which followed da Costa's work abound. One interest case was investigated in Ref. \cite{atanasov}, where a non-interacting two-dimensional electron gas(2DEG) constrained to move on a helicoid was considered. In this twisted geometry, a reminiscent of the Hall effect was observed. This fact was also pointed out in the case of carriers in graphene shaped in the same way \cite{PhysRevB.92.205425}.  Such a geometry is an object of interesting investigation in the context of the twisted bilayer graphene \cite{PhysRevB.93.035452,PhysRevB.87.121402,PhysRevB.97.125136,RIBEIRO2020100045}. This structure can be viewed as one of the fundamental pieces for more complex assemblies \cite{PhysRevB.98.155135}.

In this paper, we propose yet a conceptual design of such a QHE based on a twisted geometry. Although our finds apply to a generic non-relativistic quantum particle constrained to move on a twisted structure, we believe it might be implemented in the context of electronic membranes built from elastic materials like graphene and related materials \cite{membrane}. Also, large areas of semiconductive thin films can be developed, which in turn opens the possibility of potential future applications in the context of flexible electronics \cite{flex}. One of them can be in the context of QHE's. Here, we discuss the possibility of having a Quantum Otto engine driven by the adiabatic transformation of the number of complete twists (i.e., $2\pi$-turns) per unit length of a helicoid. The deformation we are considering is that similar to a spring, which is either compressed or stretched from its resting position. 

\section{Schr\"{o}dinger equation for a particle on a
	curved surface}
\label{EM}

In the continuum, the investigation of a non-interacting curved 2DEG is based on da Costa's approach \cite{PRA1981}. In it, the Schr\"{o}dinger equation of a free quantum particle constrained to move on an infinitely thin curved interface of the ordinary three-dimensional space was derived. It was shown that the Schr\"{o}dinger equation splits into a normal and a tangent part.  By separating these modes, a geometric induced quantum potential, which is given in terms of both the Gaussian and the mean curvatures, arises. The total wave function $\psi$ splits into its normal ($N$) and surface ($S$) components,
$\chi(q_{1},q_{2},q_{3},t)=\chi_{S}(q_1,q_{2},t) \chi_{N}(q_{3},t)$.
As a result, the normal component reads as
\begin{equation}
	i\hbar\frac{\partial}{\partial t}\chi_{N}=
	\left[
	-\frac{\hbar^2}{2m}\partial_{3}\partial^{3}
	+V_{\lambda}(q_{3})
	\right] \chi_{N},
	\label{eq:normal}
\end{equation}
where $V_{\lambda}(q_{3})$ is the transverse potential, with $\lambda$ being the squeezing parameter \cite{PRA1981},
On the other hand, the surface components are
\begin{align}
	i\hbar\frac{\partial}{\partial t}\chi_{S}= {}
	&
	\Bigg[\frac{-\hbar^2}{2m}\frac{1}{\sqrt{g}}\partial_{a}
	\left(\sqrt{g}g^{ab}\partial_{b}\right)
	+V_{S}\Bigg] \chi_{S},
	\label{eq:surface}
\end{align}
where $g^{ab}$ are the contravariant components of the metric tensor of the surface, $g=\det(g^{ab})$, $a,b=1,2$ and
$V_{S}(q_{1},q_{2})$ is the geometric induced quantum potential, which is written as 
\begin{equation}
V_{S}(q_1,q_2)=-\frac{\hbar^2}{2m}(\rm{M}^{2}-\rm{K}_G),
\label{eq:ptg}
\end{equation}
where $\rm{M}$ is the mean curvature and $\rm{K}_G$ is the
Gaussian curvature of the surface. In terms of principal curvatures, $\kappa_1$, and $\kappa_2$, at a given point on the surface, they are given by $\rm{M}=-(\kappa_1+\kappa_2)/2$ and $\rm{K}_G=\kappa_1\kappa_2$.

In the next section, we derive the Schr\"{o}dinger equation for this case. We consider a 2DEG confined in an infinite helical stripe, and find the first few energy levels of the system. 

\section{A quantum particle on a helicoid}

A helicoid can be parametrized by the following set of equations \cite{geometry}:
\begin{align}
x &= \rho\cos(\omega z),  \nonumber \\
y &= \rho\sin(\omega z), \nonumber \\
z &= z, 
\end{align}
where $\omega=2\pi S$, with $S$ being the number of complete twists (i.e., $2\pi$-turns) per unit length of the helicoid. $\rho$ is the transverse direction across the helicoid, starting from the $z$-axis.  The infinitesimal line element on the helicoid is given by
\begin{equation} 
	ds^2 = d\rho^2 + (1 + \omega^2\rho^2)dz^2\;,
\end{equation}
where we have used the coordinates $(z, \rho)$ to characterize a point on the helicoid. The metric components are thus obtained as
\begin{align}
	g_{\rho \rho} &= 1, \nonumber \\
	g_{zz} &= 1 + \omega^2\rho^2, \nonumber \\
	g_{z \rho} &= 0, \nonumber \\
	g_{\rho z} &= 0. \label{E:metriczheli}
\end{align}
The square root of the determinant of the metric is given by $\sqrt{g} = \sqrt{1 + \omega^2\rho^2}$, and the principal curvatures, $\kappa_1$ and $\kappa_2$, are given by
\begin{eqnarray}
\kappa_1 &=& \frac{\omega}{1+\omega^2\rho^2}, \nonumber \\
\kappa_2 &=& -\frac{\omega}{1+\omega^2\rho^2}. \label{helic}
\end{eqnarray}
From there, we observe that a helicoid is a minimal surface, since the mean curvature ${\rm M}$ vanishes at any given point on it, that is,
\begin{equation} \label{E:meancurvatureheli}
	{\rm M} \equiv \frac{1}{2}(\kappa_1 + \kappa_2) = 0.
\end{equation}
The Gaussian curvature ${\rm K_G}$ is found to be
\begin{equation} \label{E:gaussiancurvatureheli}
	{\rm K_G} = \kappa_1\kappa_2 = -\frac{\omega^2}{(1+\omega^2\rho^2)^2}.
\end{equation}
The geometric quantum potential $V_S$ is given by
\begin{equation} \label{E:curvpotentialheli}
	V_S = -\frac{\hbar^2}{2m}\left({\rm M}^2 - {\rm K_G} \right) =  -\frac{\hbar^2}{2m}\frac{\omega^2}{(1+\omega^2\rho^2)^2}.
\end{equation}
Therefore, the Schr\"{o}dinger equation for a particle on a helicoid can be written as 
\begin{align} \label{E:hamiltoniancurv}
	i\hbar \partial_t \chi_{S}&=-\frac{\hbar^2}{2m}\left[\frac{1}{a} \left(\partial_z\left( \frac{1}{a}\partial_z\chi_S \right) + \partial_{\rho}(a\partial_{\rho}\chi_S ) \right)  \right]\notag \\& - \frac{\hbar^2}{2m}\frac{\omega^2}{(1+\omega^2\rho^2)^2}\chi_s,
\end{align}
with $a\equiv \sqrt{1 + \omega^2\rho^2}$. Next, we take the ansatz 
\begin{equation}
	\chi_{S}(\rho,z)=e^{il\omega z}f\left(\rho\right),
\end{equation}
with $l \in \mathbb{N}$, which is due to the quantized angular momentum along $z$ axis. This way, $\omega z$ will be the azimuthal angle around the $z$ axis. Considering that the wave function has to be normalized with respect to the infinitesimal area $d\rho dz$, we make the substitution $\chi_S \to (1/\sqrt{a})\chi_S$ in Eq. (\ref{E:hamiltoniancurv}). Only terms involving derivatives with respect to $\rho$ are affected \cite{atanasov}.  Omitting the algebra, we rewrite Eq. (\ref{E:hamiltoniancurv}) as
\begin{align} 
	E\chi_{s}=-\frac{\hbar^2}{2m}\frac{\partial^{2}\chi_{s}}{\partial\rho^{2}}+\frac{\hbar^2}{2m(1+\omega^2\rho^2)}\frac{\partial^2 \chi_{s}}{\partial z^2}+V_{eff}(\rho)\chi_{s}, \label{scho}
\end{align}
where 
\begin{equation}
	V_{eff}(\rho)=-\frac{\hbar^2\omega^2}{4m(1+\omega^{2}\rho^2)^{2}}\left(1+\frac{\omega^{2}\rho^2}{2}\right)\label{effective}\;,
\end{equation}
is the effective quantum potential induced by the geometry of a helical surface. By defining the quantities $\xi\equiv\omega\rho$ and $\varepsilon\equiv 2mE/\hbar^2\omega^2$, Eq. (\ref{scho}) is rewritten as
\begin{align} \label{E:hamiltoniancurvnorm}
	\varepsilon\chi_{s}=-\frac{d^{2}\chi_{s}}{d\xi^{2}}+\Bigg[\frac{l^{2}}{1+\xi^{2}}-\frac{1}{2(1+\xi^{2})^{2}}\left(1+\frac{\xi^{2}}{2}\right)\Bigg]\chi_{s}.
\end{align}

Now, we address the problem of a quantum particle confined in the radial direction, $\rho$. For simplicity, we will consider $\rho_1=0$ ($\xi_1=0$) e $\rho_2=\rho_L$ ($\xi_2=\rho_L\omega$). 
With the use of the Maple software, we obtain the following solution for Eq. (\ref{E:hamiltoniancurvnorm}) \cite{Book.Ronveaux.1995}:
\begin{figure}[!b!]
	\centering
	\includegraphics[scale=0.35]{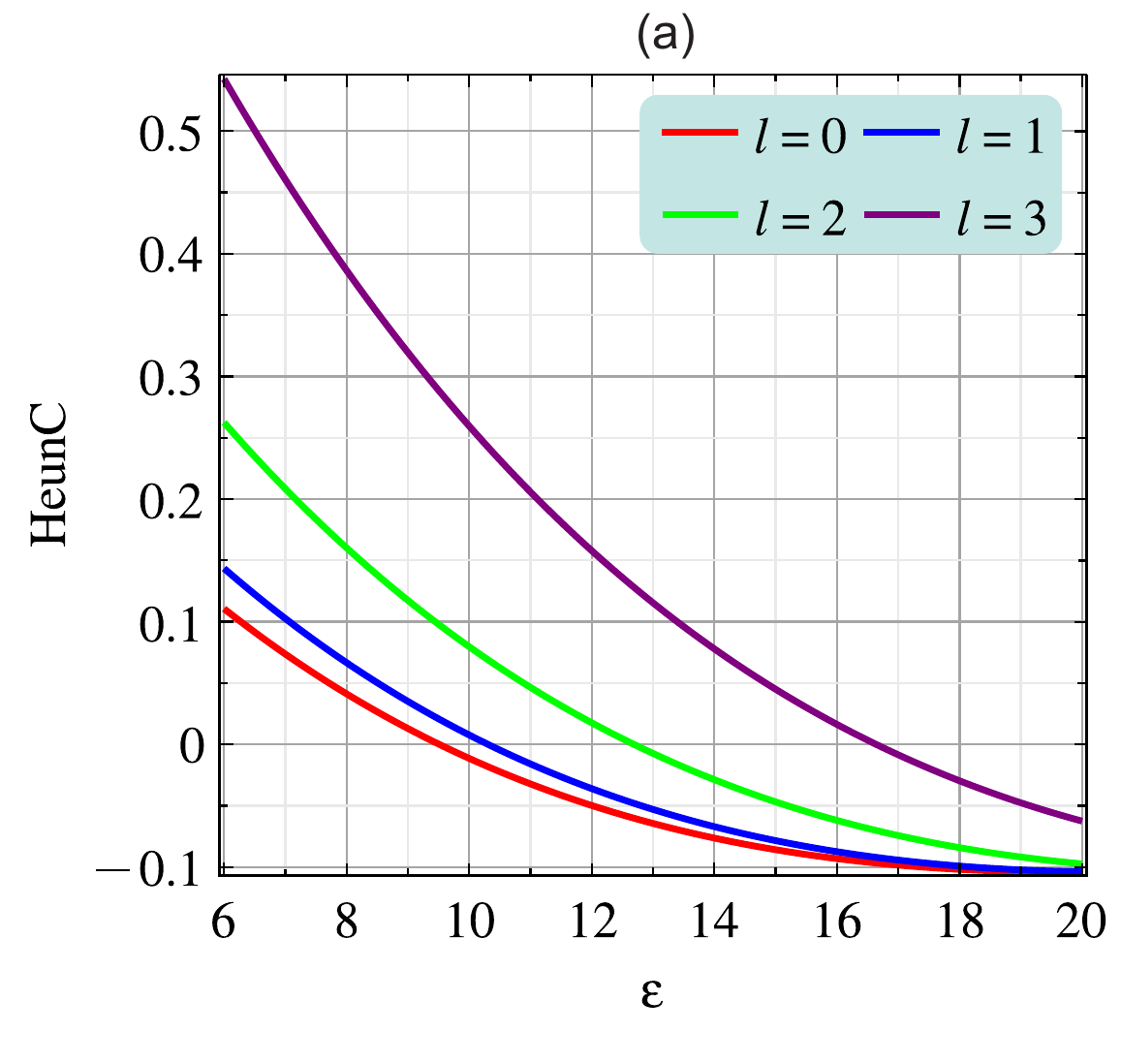}
	\includegraphics[scale=0.35]{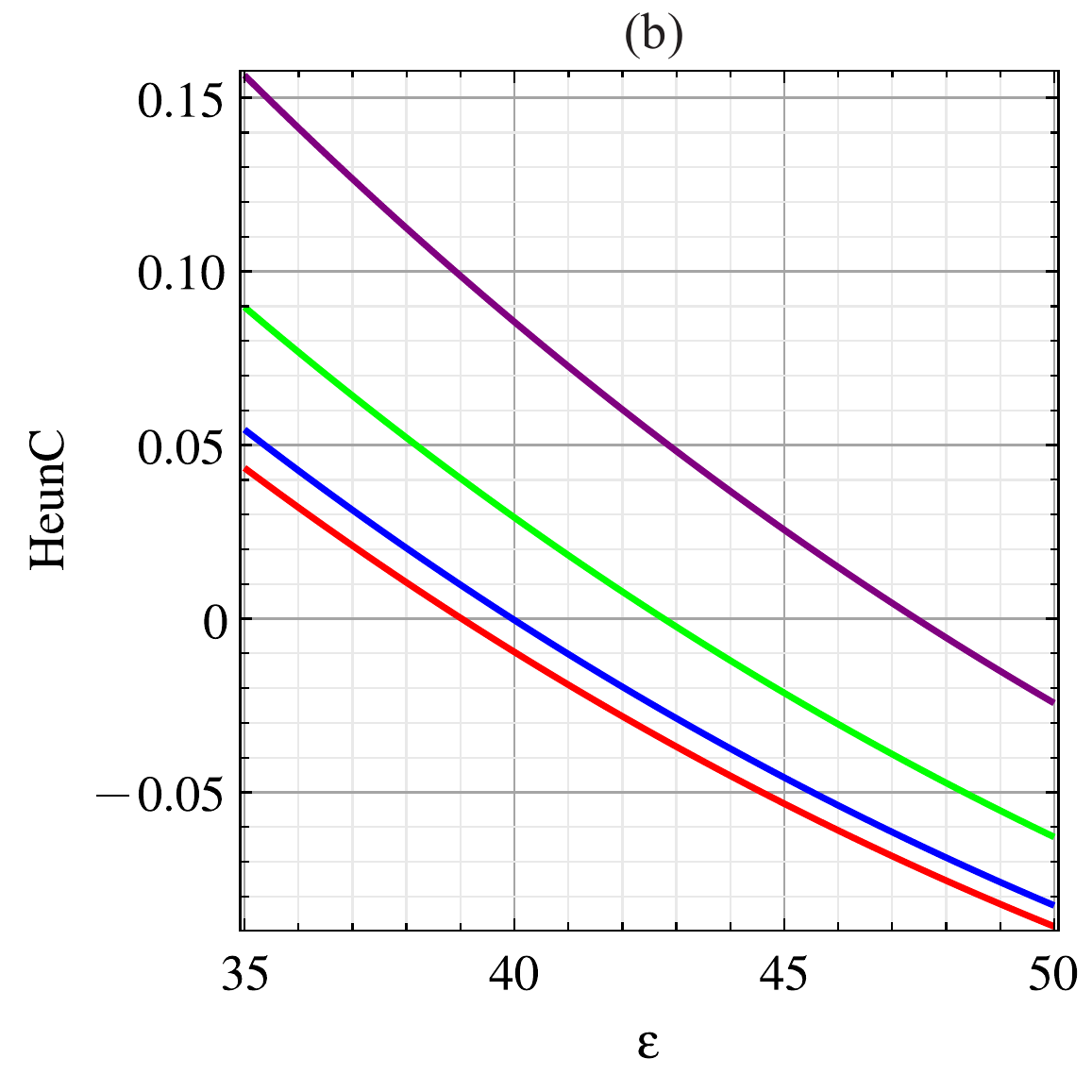}
	\caption{\small Plot of the HeunC function versus energy. The eigenvalues for a particle in a helical stripe can be found where it reaches zero values. The first few levels are depicted in (a) for $\omega\rho_2=1$ and in (b) for $\omega\rho_2=0.5$.}\label{cat}
\end{figure}
\begin{align}
	\chi_S \left( \xi \right)&= \left( {1+\xi}^{2} \right)^{\frac{1}{2}+\frac{\sqrt 5}{4}} \notag \\ &\times\Bigg[{\it C_1}{\rm HeunC} \left( 0,-\frac{1}{2},\frac{\sqrt {5}}{2},-\frac{\varepsilon}{4},
	\frac{5}{8}-\frac{l^2}{4}+\frac{\varepsilon}{4},-{\xi}^{2} \right) \nonumber\\ &+ {\it C_2}\,\xi\,{\rm HeunC}
	\left( 0,\frac{1}{2},\frac{\sqrt {5}}{2},-\frac{\varepsilon}{4},\frac{5}{8}-\frac{l^2}{4}+
	\frac{\varepsilon}{4},-{\xi}^{2} \right) \Bigg],\label{sol}
\end{align}
where the ${\rm HeunC}(\alpha,\beta,\gamma,\delta,\eta,z)$ function is a local (Frobenius) solution to Heun's Confluent equation, and $C_{1}$ and $C_{2}$ are arbitrary constants. Considering the boundary condition $\chi_S(0)=0$, it yields
\begin{equation}
	\chi_S(0)=C_1=0\;,
\end{equation}
since ${\rm HeunC}(\alpha,\beta,\gamma,\delta,\eta,0)=1$. 
In this way, the wavefunction for this problem is given by
\begin{equation} 
	\chi_S \left( \xi \right)={\it C_2}\,\xi\,{\rm HeunC}
	\left( 0,\frac{1}{2},\frac{\sqrt {5}}{2},-\frac{\varepsilon}{4},\frac{5}{8}-\frac{l^2}{4}+
	\frac{\varepsilon}{4},-{\xi}^{2} \right).\label{solbox}
\end{equation}
Taking into account that the wavefunctions also vanish at $\xi=\xi_2$, that is, $\chi(\xi_2)=0$, the energy levels of the system can be found from the plot of the ${\rm HeunC}$ function versus $\varepsilon$. Obviously, they are found where the values of it are null (see Fig. \ref{cat}). 

In the next section, we describe the quantum Otto cycle considering an adiabatic transformation of the radial size of a helicoid as well as the number of complete twists (i.e., $2\pi$-turns) per unit length of it. 

\section{\label{sec:level2}Quantum Otto cycle for a 2DEG in a helicoid}

\begin{figure}[!h]
\centering
\includegraphics[scale=0.35]{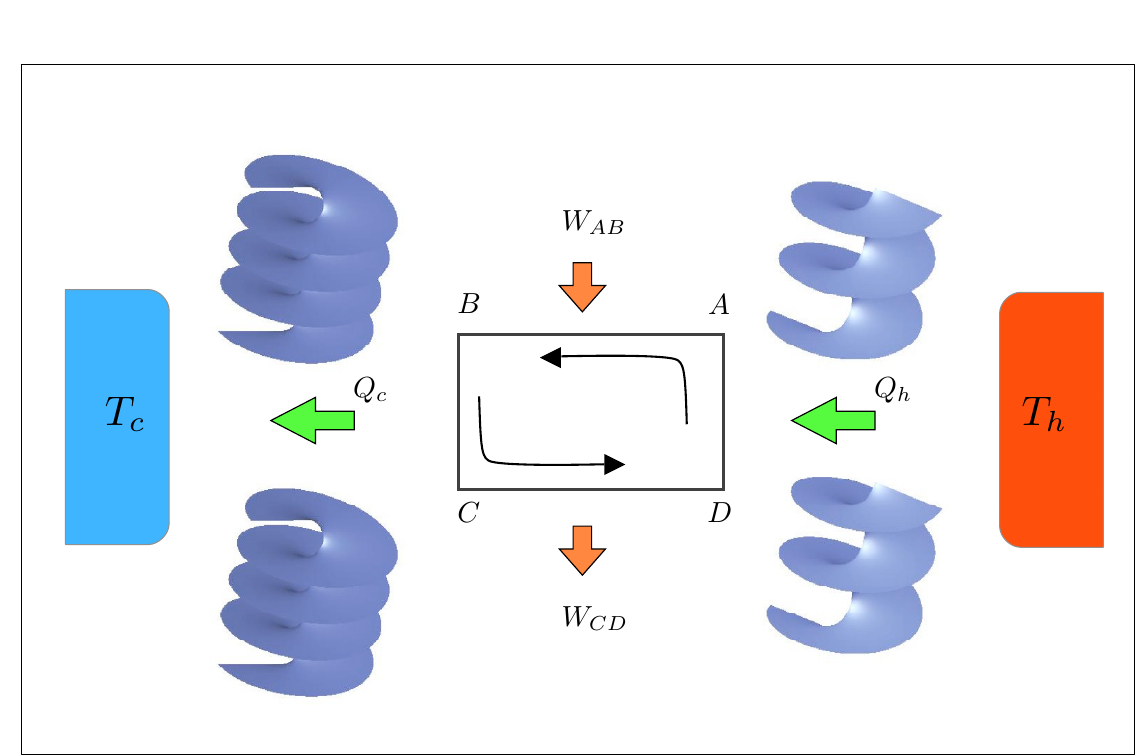}
\caption{A Quantum Otto Cycle for a 2DEG on an infinity helical stripe. The adiabatic strokes consist of modifications in the transverse direction across the helicoid and in the number of complete twists per unit length. The paths AB and CD are adiabatic strokes while BC and DA are isochoric strokes.} \label{helicoicycle}
\end{figure}

The Otto cycle consists in four strokes that connect
different states of the system, A, B, C, and D. It is depicted in Fig. \ref{helicoicycle}. At A, a 2DEG  is confined in an infinity helical stripe with a given number of complete twists  per unit length, $S_h$. At this point, the system is at thermal equilibrium with the hot bath at the temperature $T_h$. This way, the population of the $n^{th}$-level is 
\begin{equation}
P_{n,A} = Z^{-1}_{h}e^{-\frac{E_{n,h}}{k_B T_h}},
\end{equation}
where $E_{n,h}$ is the $n^{th}$-level eigenenergy at $A$ and 
\begin{equation}
Z_h=\sum_n e^{-\frac{E_{n,h}}{k_B T_h}}.
\end{equation}
The dimensionless size of the system is $\xi_h$. 
Next, the system is decoupled from the hot bath and the number of complete twists is adiabatically modified until it reaches a value $S_c$ at B. The size of the system changes to $\xi_c$ (in principle, $S$ and $\rho$ have been modified). By keeping the process adiabatic, it ensures that the level of populations do not change. Then, $P_{n,A} = P_{n,B}$. The change in the energy of the system can be attributed solely to work, which we call $W_{AB}$. In the second stroke, the system is coupled to a cold thermal bath at temperature $T_c$ and it reaches thermal equilibrium at C. Thus, 
\begin{equation}
P_{n,C} = Z^{-1}_{c}e^{-\frac{E_{n,c}}{k_B T_c}}, 
\end{equation}
where $E_{n,c}$ is the $n^{th}$-level eigenenergy at $C$ and 
\begin{equation}
Z_c=\sum_n e^{\frac{-E_{n,c}}{k_B T_c}}.
\end{equation}
The change in the energy of the system can be attributed to heat
exchange with the cold bath $Q_c$. In the third stroke, the system is decoupled from the cold bath and the number of complete twists  per unit length is adiabatically modified until it returns to $S_h$ and the size of the system to $\xi_h$ at D. The level populations do not change, so we have $P_{n,D} = P_{n,C}$. Therefore, all the energy exchanged is work, which we call $W_{CD}$. In the final stroke, the system is coupled to the hot bath, ending at thermal equilibrium with it at A. The thermodynamic cycle is then, closed. The size of the system as well as its number of complete twists are kept constant and the exchanged energy is heat with the hot bath, $Q_h$.

The heat exchanged with the baths is given by the energy difference between the initial and the final states of 
the isochoric strokes.  The heat exchanged with the hot bath is given by 
\begin{equation}\label{eq6}
	Q_h=\sum_{n}E_n^h(p_n^h-p_n^c),
\end{equation}
while the heat exchanged with the cold one is
\begin{equation}\label{eq7}
	Q_c=\sum_{n}E_n^c(p_n^c-p_n^h).
\end{equation}
These definitions imply $Q>0$ ($Q<0$) meaning that heat is absorbed (released) from (to) the heat reservoirs, respectively. Therefore, the total work $W$ produced by the heat engine in the adiabatic is, by energy conservation, the excess heat
\begin{equation}\label{work1}
	W=Q_h+Q_c=\sum_{n}(E_n^h-E_n^c)(p_n^h-p_n^c).
\end{equation}
We shall consider a two-level system, with  $E_1\equiv E_{g,i}$ and $E_{2,i}\equiv E_e$,  with $i=h,c$. Equations (\ref{eq6}) and (\ref{eq7}) are written as

\onecolumngrid
\begin{equation}
	Q_{c}=-E_{1}^{c}\left[\frac{e^{-\frac{E_{gh}}{k_{B}T_{h}}}}{\left(e^{-\frac{E_{g,h}}{k_{B}T_{h}}}+e^{-\frac{E_{e,h}}{k_{B}T_{h}}}\right)}-\frac{e^{-\frac{E_{gc}}{k_{B}T_{c}}}}{\left(e^{-\frac{E_{g,c}}{k_{B}T_{c}}}+e^{-\frac{E_{e,c}}{k_{B}T_{c}}}\right)}\right]-E_{2}^{c}\left[\frac{e^{-\frac{E_{eh}}{k_{B}T_{h}}}}{\left(e^{-\frac{E_{g,h}}{k_{B}T_{h}}}+e^{-\frac{E_{e,h}}{k_{B}T_{h}}}\right)}-\frac{e^{-\frac{E_{ec}}{k_{B}T_{c}}}}{\left(e^{-\frac{E_{g,c}}{k_{B}T_{c}}}+e^{-\frac{E_{e,c}}{k_{B}T_{c}}}\right)}\right]\label{cold}
\end{equation}
and 
\begin{equation}
	Q_{h}=E_{1}^{h}\left[\frac{e^{-\frac{E_{gh}}{k_{B}T_{h}}}}{\left(e^{-\frac{E_{g,h}}{k_{B}T_{h}}}+e^{-\frac{E_{e,h}}{k_{B}T_{h}}}\right)}-\frac{e^{-\frac{E_{gc}}{k_{B}T_{c}}}}{\left(e^{-\frac{E_{g,c}}{k_{B}T_{c}}}+e^{-\frac{E_{e,c}}{k_{B}T_{c}}}\right)}\right]+E_{2}^{h}\left[\frac{e^{-\frac{E_{eh}}{k_{B}T_{h}}}}{\left(e^{-\frac{E_{g,h}}{k_{B}T_{h}}}+e^{-\frac{E_{e,h}}{k_{B}T_{h}}}\right)}-\frac{e^{-\frac{E_{ec}}{k_{B}T_{c}}}}{\left(e^{-\frac{E_{g,c}}{k_{B}T_{c}}}+e^{-\frac{E_{e,c}}{k_{B}T_{c}}}\right)}\right]\label{hot}.
\end{equation}
\twocolumngrid
By defining $\delta_c\equiv\epsilon_{e,c}-\epsilon_{g,c}$,  $\delta_h\equiv\epsilon_{e,h}-\epsilon_{g,h}$ and considering $\omega_{h}=\sqrt{\alpha}\omega_{c}$, together with the condition
\begin{equation}
	\frac{\Delta_{c}}{\Delta_{h}}=\frac{\frac{\hbar^{2}\omega_{c}^{2}}{2m}\delta_c}{\frac{\hbar^{2}\omega_{h}^{2}}{2m}\delta_h}<1,\label{condition}
\end{equation}
the following bound must be satisfied in what follows:
\begin{equation}
	\alpha>\frac{\delta_{c}}{\delta_{h}}.\label{bound1}
\end{equation}
First, we consider the transition from $\xi_c\equiv\omega_c\rho_c=0.5$ to $\xi_h\equiv\omega_h\rho_h=1$. In this case, we obtain $\delta_c\approx 0.93$ and $\delta_h\approx 0.79$. The bound (\ref{bound1}) is then $\alpha\gtrapprox1.18$. 
Defining $\widetilde{Q}_{c}\equiv Q_{c}/ \left(\frac{\hbar^{2}}{2m\rho_{c}^{2}}\right)$, $\widetilde{Q}_{h}\equiv Q_{h}/\left(\frac{\hbar^{2}}{2m\rho_{c}^{2}}\right)$ and $\widetilde{W}\equiv W/\left(\frac{\hbar^{2}}{2m\rho_{c}^{2}}\right)$, together with the dimensionless parameters 
$\varsigma=\hbar^{2}/2Mk_{B}\rho_{c}^{2}T_{c}$, $\theta= T_{h}/T_{c}$, Eqs. (\ref{cold}) and (\ref{hot}) are rewritten, respectively, as
\begin{equation}
	\widetilde{Q}_{c}=-\frac{\delta _{c}\left[ e^{-\varsigma \left( 0.5^{2}\epsilon
			_{gc}+\frac{\epsilon _{eh}}{\theta }\right) }-e^{-\varsigma \left( \frac{
				\epsilon _{gh}}{\theta }r^{2}+0.5^{2}\epsilon _{ec}\right) }\right] }{
		4\left( e^{-0.5^{2}\varsigma \epsilon _{gc}}+e^{-0.5^{2}\varsigma \epsilon
			_{ec}}\right) \left( e^{-\frac{\varsigma \epsilon _{gh}}{\theta }r^{2}}+e^{-
			\frac{\varsigma \epsilon _{eh}}{\theta }r^{2}}\right) }
\end{equation}
and 
\begin{equation}
	\widetilde{Q}_{h}=\frac{\delta _{h}\left[ e^{-\varsigma \left( 0.5^{2}\epsilon
			_{gc}+\frac{\epsilon _{eh}}{\theta }r^{2}\right) }-e^{-\varsigma \left(
			0.5^{2}\epsilon _{ec}+\frac{\epsilon _{gh}}{\theta }r^{2}\right) }\right] }{
		\left( e^{-0.5^{2}\varsigma \epsilon _{gc}}+e^{-0.5^{2}\varsigma \epsilon
			_{ec}}\right) \left( e^{-\frac{\varsigma \epsilon _{gh}}{\theta }r^{2}}+e^{-
			\frac{\varsigma \epsilon _{eh}}{\theta }r^{2}}\right) }r^{2}.
\end{equation}
We also consider an opposite  transition, from $\omega_c\rho_c=1$ to $\omega_h\rho_h=0.5$. For this case, $\delta_c\approx 0.79$ and $\delta_h\approx 0.93$ and $\alpha\gtrapprox 0.85$. 
The heat exchanged with the cold bath is now given by  
\begin{equation}
	\widetilde{Q}_{c}=-\frac{\delta _{c}\left[ e^{-\varsigma\left(\epsilon
			_{gc}+\frac{0.5^{2}\epsilon _{eh}}{\theta }r^{2}\right) }-e^{-\varsigma \left(
			\epsilon _{ec}+\frac{0.5^{2}\epsilon _{gh}}{\theta }r^{2}\right) }\right] }{
		\left( e^{-\varsigma \epsilon _{gc}}+e^{-\varsigma \epsilon _{ec}}\right) \left( e^{-
			\frac{0.5^{2}\varsigma \epsilon _{gh}}{\theta }r^{2}}+e^{-\frac{0.5^{2}\varsigma
				\epsilon _{eh}}{\theta }r^{2}}\right) },  \label{cold2}
\end{equation}
and with the hot one is
\begin{equation} 
	\widetilde{Q}_{h}=\frac{\delta _{h}\left[ e^{-\varsigma\left(\epsilon
			_{gc}+\frac{0.5^{2}\epsilon _{eh}}{\theta }r^{2}\right) }-e^{-\varsigma \left(
			\epsilon _{ec}+\frac{0.5^{2}\epsilon _{gh}}{\theta }r^{2}\right) }\right] }{
		4\left( e^{-\varsigma \epsilon _{gc}}+e^{-\varsigma \epsilon _{ec}}\right) \left(
		e^{-\frac{0.5^{2}\varsigma \epsilon _{gh}}{\theta }r^{2}}+e^{-\frac{0.5^{2}\varsigma
				\epsilon _{eh}}{\theta }r^{2}}\right) }r^{2}.  \label{hot2}
\end{equation}
Finally, we take into account the limit $\rho>>1$ in Eq. (\ref{E:hamiltoniancurvnorm}). In this case, we obtain the differential equation 
\begin{equation}
	\frac{d^{2}\chi_{s}(\xi)}{d\xi^{2}}+\frac{(l^{2}-1/4)\chi_{s}(\xi)}{\xi^{2}}=\epsilon \chi_{s}(\xi),
\end{equation}
\begin{figure}[!t]
	\centering
	\includegraphics[width=0.45\textwidth]{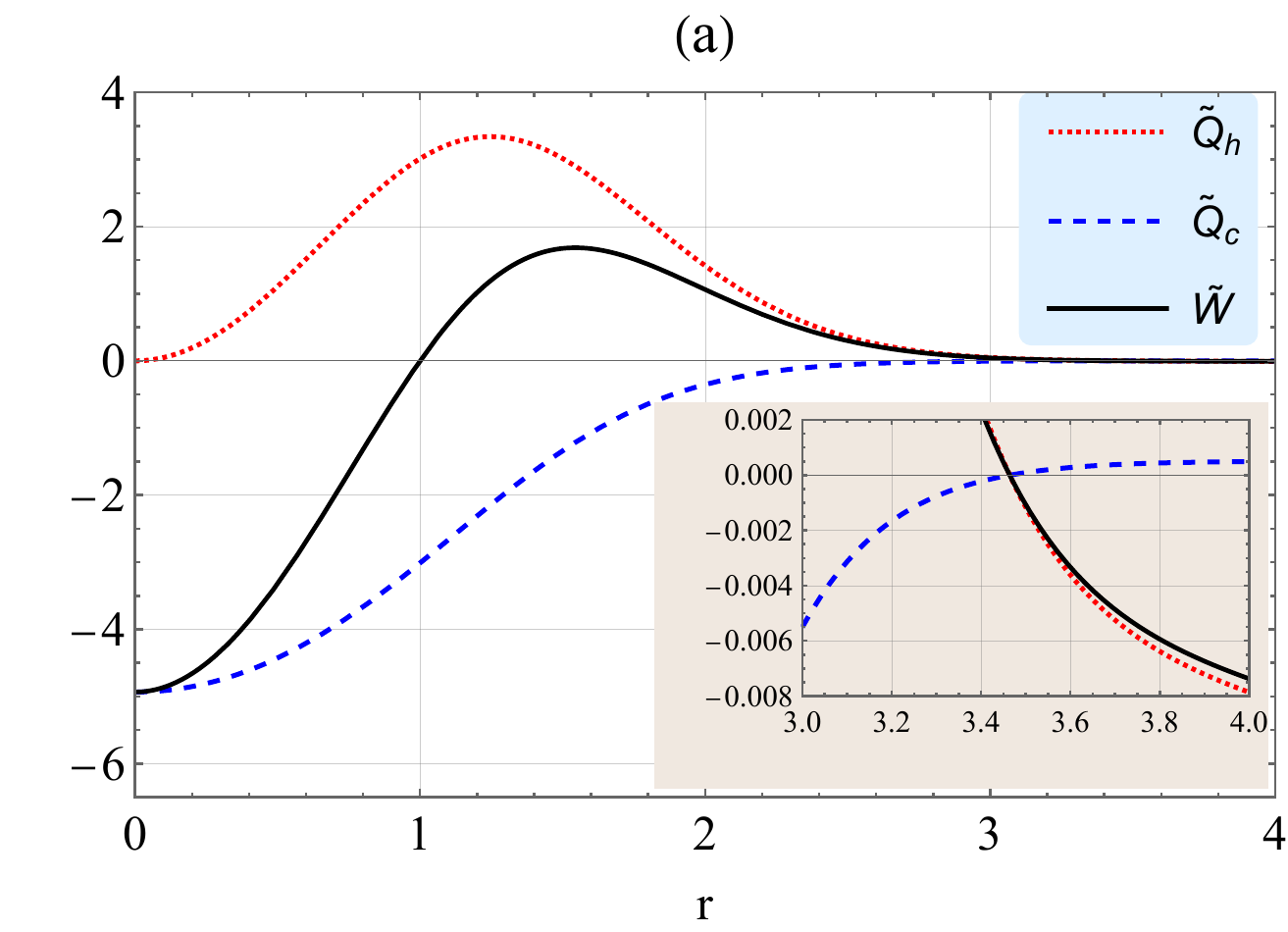}
	\includegraphics[width=0.45\textwidth]{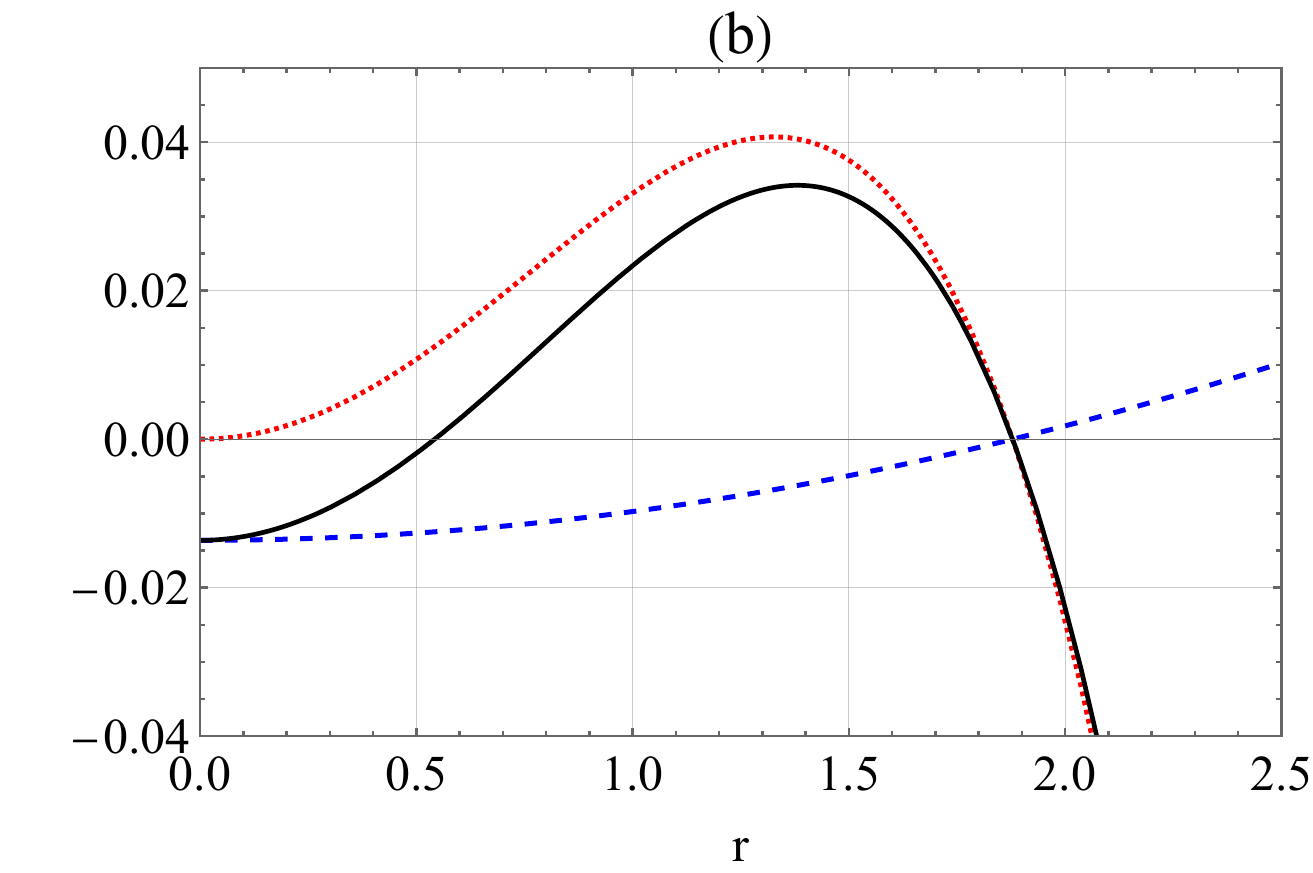}
	\includegraphics[width=0.45\textwidth]{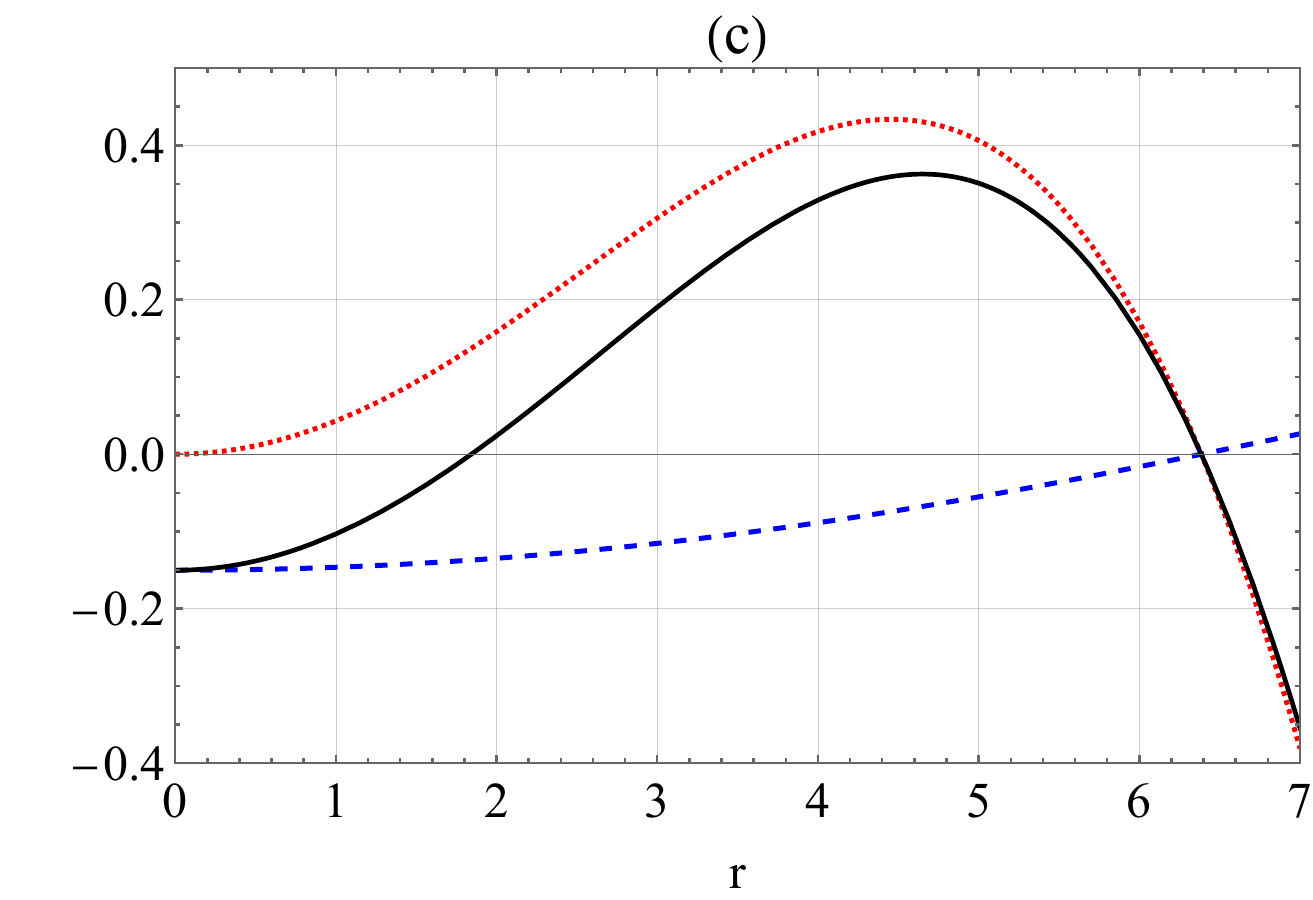}
	\caption{\small The heat exchanges of the working substance with the hot and cold reservoirs and the work done are depicted. The flat case is shown in (a), for which $W>0$ for $1<r\lessapprox3.464$. For $r<1$, the system works as a heater and for $r\gtrapprox3.464$, it behaves as a refrigerator, similarly to a classical Otto machine. In (b), $\xi_c\equiv\omega_c\rho_c=0.5$ and $\xi_h\equiv\omega_h\rho_h=1$; $W>0$ for  $0.54\lessapprox r\lessapprox1.878$. The curves shift to lower values of $r$ in comparison to the flat case. In this case, the transverse direction $\rho$ may be kept constant ($r=1$).  In (c), $\xi_c\equiv\omega_c\rho_c=1$ and $\xi_h\equiv\omega_h\rho_h=0.5$; $W>0$ for $1.844\lessapprox r\lessapprox6.386$ and the curves shift to higher values of $r$ in comparison to the flat case. We have set $T_h\equiv12T_c$ and $\hbar^{2}/{2Mk_{B}\rho_{c}^{2}T_{c}}\equiv1$. }\label{figW}
\end{figure}
\begin{figure}[!t]
	\centering
	\includegraphics[width=0.45\textwidth]{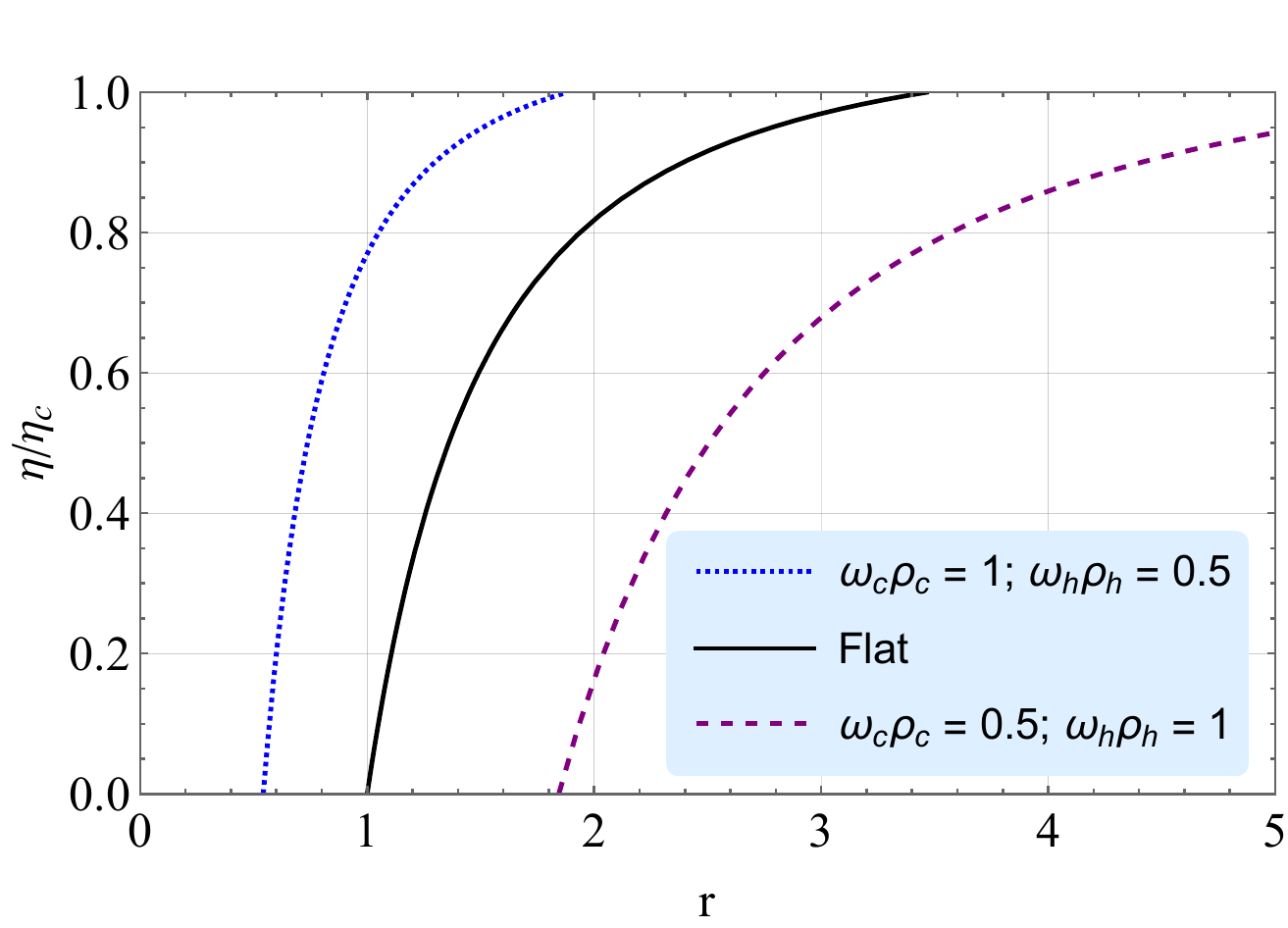}
	\caption {The efficiency $\eta/(1-T_c/T_h)$ against the compression ratio $r=\rho_c/\rho_h$ is depicted. The temperature ratio is set to $T_h/T_c=12$. For a flat sample, only the transverse direction $\rho$ can be altered and it behaves similarly to a classical Otto engine. The curvature shifts the efficiency curves to either lower or higher values of $r$, now allowing such transverse direction to be kept unaltered ($r=1$).} 
	\label{eff}
\end{figure}
\begin{figure}[!t]
	\centering
	\includegraphics[width=0.45\textwidth]{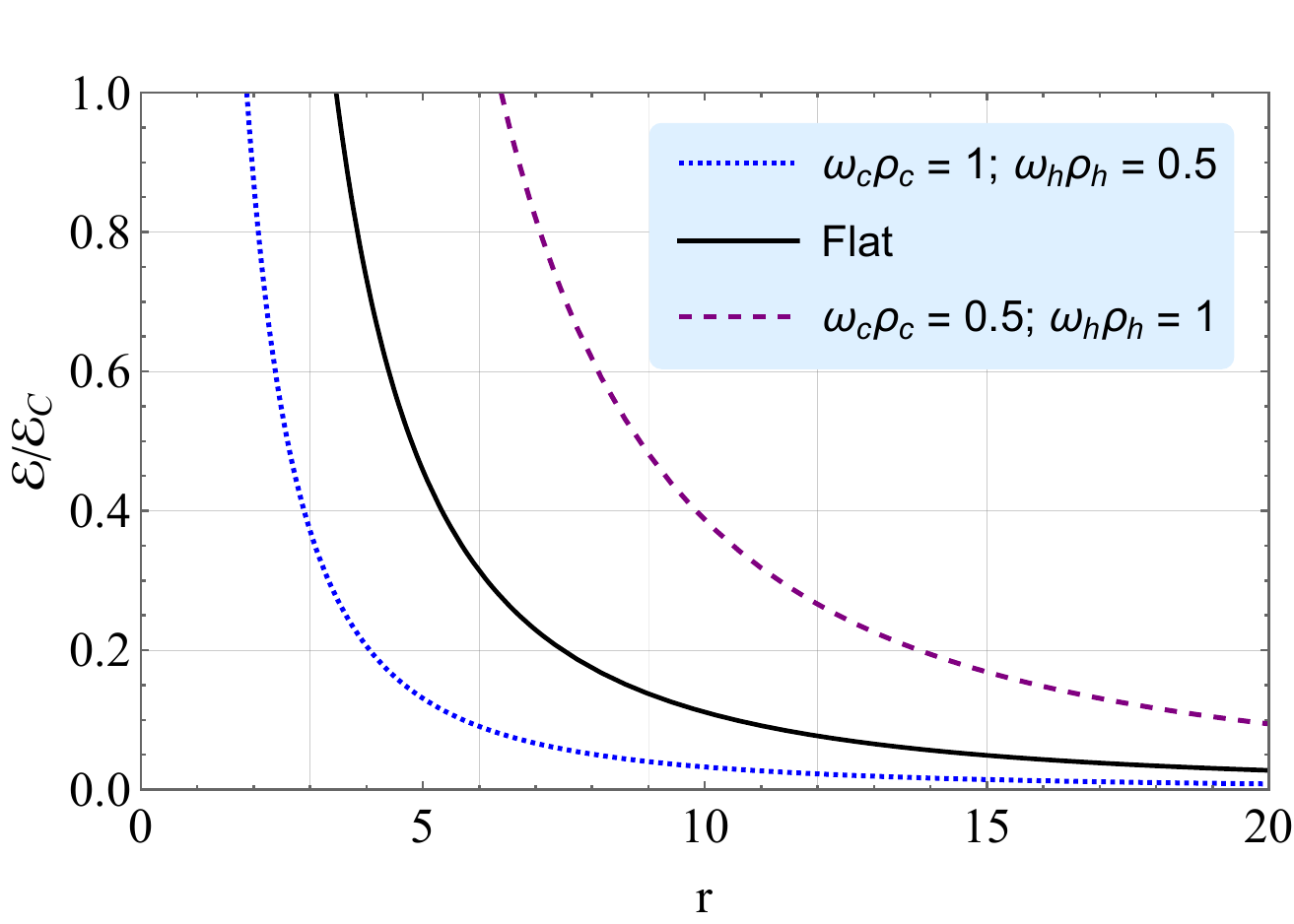}
	\caption {The coefficient of performance $\varepsilon/[T_c/(T_h-T_c)]$ against the compression ratio $r=\rho_c/\rho_h$ is depicted. The temperature ratio is set to $T_h/T_c=12$. The refrigerator performance can be altered/improved by curvature effects due to a twist geometry.} \label{cof}
\end{figure}
whose solution is given by
\begin{equation}
	\chi_{s}(\xi)=c_{1}\sqrt{\xi} J_{l}\left(\sqrt{\epsilon}\,\xi \right)+c_{2}\sqrt{\xi}\,Y_{l}\left(\sqrt{\epsilon}\,\xi \right),
\end{equation}
where $J_{l}(\sqrt{\epsilon}\,\xi)$ is the Bessel function of the first kind and $Y_{l}(\sqrt{\epsilon}\,\xi)$ is the Bessel function of the second kind.
Putting $c_2\equiv0$, we have 
\begin{equation}
	\chi_{s}(\xi)=c_{1}\sqrt{\xi} J_{l}(\sqrt{\epsilon}\,\xi).\label{yn}
\end{equation}
Considering the asymptotic behaviour of (\ref{yn}), $\xi>>1$, and substituting back $\xi\equiv\omega\rho$ and $\epsilon\equiv 2mE/{\omega^{2}\hbar^{2}}$, we find
\begin{equation}
	\chi(\rho)=c_{1}\sqrt{\frac{2\pi}{\sqrt{\frac{2mE}{\omega^{2}\hbar^{2}}}\omega\rho}}\cos\left(\sqrt{\frac{2mE}{\omega^{2}\hbar^{2}}}\omega\rho-\frac{l\pi}{2}-\frac{\pi}{4}\right).\label{ynn}
\end{equation}
Applying the boundary condition $\chi(\rho_B)=0$ to (\ref{ynn}), it yields 
\begin{equation}
	E\equiv E_{n,l}=\frac{\hbar^{2}}{2m\rho_{B}^{2}}\left[\left(n+\frac{1}{2}\right)\pi+\frac{l\pi}{2}+\frac{\pi}{4}\right]^{2}.
\end{equation}
We will refer to this case as a {\it flat} one since the influence of the curvature is lost, similarly, if we have considered a flat sample at the beginning. The expression for $\widetilde{Q}_{c}$, $\widetilde{Q}_{h}$ and $\widetilde{W}$ are easy to obtain as before but we will omit here. Nevertheless, their plots are depicted and discussed in the next section. 
\section{Results and discussion}

We start by analyzing the heat exchanged $Q_h$ ($Q_c$) with the hot (cold) reservoir and the work done $W$ in the function of the compression ratio $r=\rho_c/{\rho_h}$. We set $\varsigma\equiv1$ and $\theta\equiv12$. In Fig. \ref{figW}, they are depicted as a function of $r$. By changing the radius of a helicoid we are changing its area (or volume, if we consider that the sample has a fixed thickness). We also consider, while doing so, an adiabatic modification of the number of complete twists per unit length, here represented by $\omega$. First of all, for a flat sample (solid line), we need $r>1$ to achieve positive work, which is reasonable because in this regime the energy gaps are squeezed when the system is in contact with the cold reservoir and they are expanded when the system is in contact with the hot one (see inequality (\ref{condition})). 
Notice that we can not increase the radius $\rho_c$ indefinitely since the positive work condition is lost. The machine behaves as an engine in the interval $1<r\lessapprox3.464$.
Furthermore, for $r\gtrapprox3.464$, the signs of the heat exchanged are inverted and the system starts to withdraw heat from the cold reservoir and transfer heat into the hot reservoir. This means that the system behaves as a refrigerator at cost of some work. For $r<1$, the system behaves as a heater. So, the flat case corresponds to a common Otto machine, with no different behavior from the classical one.  

Now, we discuss the influence of the adiabatic modification of the curvature of the sample. In this case, two parameters can be varied in order to achieve either energy level compression or expansion. One is the radial size of a helical stripe, which is represented by $r$. The other one is the parameter $\omega$, which is related, as pointed out above, to the number of complete twists per unit length of the surface. By tailoring the latter, we observe the engine operation changes to values of $r$ not expected for a flat sample (see Fig. \ref{figW}). It shows that the quantum Otto engine may extract work at any other value of $r$, including $r = 1$. This last case corresponds to a fixed radial size of the helical strip. Notice that the size of a box in the flat sample can not be kept fixed. When $\xi_c\equiv\omega_c\rho_c=0.5$ and $\xi_h\equiv\omega_h\rho_h=1$, $Q_c$ $Q_h$ and $W$ shift towards $r < 1$, which in turns allows the possibility of work extraction at $0.54\lessapprox r\lessapprox1.878$. For $\xi_c\equiv\omega_c\rho_c=1$ and $\xi_h\equiv\omega_h\rho_h=0.5$, they shift higher values of $r$, and the possibility of work extraction occurs at $1.844\lessapprox r\lessapprox6.386$. Quantum engines for any compression ratio can be achieved beyond the two-level approximation. This requires the control of the relation between the size of the stripe and the parameter $\omega$, which will ensure that all the energy levels have the appropriate values. In the same way, $\omega$ could be optimized for reaching maximum work extraction at any compression ratio or for producing maximum heat extraction, or refrigeration efficiency.

The efficiency of a heat engine is calculated by $\eta\equiv W/Q_h$. For a two-level system, it yields
\begin{equation}
	\eta=1-\frac{\Delta_{c}}{\Delta_{h}}\;.
\end{equation}
It is analyzed in Fig. \ref{eff}. We also evaluate the Coefficient of Performance (COP) for the refrigerator operation mode, which is obtained as $\varepsilon\equiv Q_c/W$. We compare the three cases in Eq. (\ref{cof}).  Notice that the COP is a monotonically decreasing function of the compression ratio $r$. As we have said above, the deformation we are considering is that for a spring, which is either compressed or stretched from its resting position. 

It is well known that the power output of an ideal engine, including the Otto one, is zero. The real ones operate in finite time, so the working substance is almost never in equilibrium with the heat baths. The question about the efficiency at maximum power output, for instance, should be addressed considering a mechanism to speed up the evolution of an arbitrary quantum dynamical system yielding a shortcut to adiabaticity \cite{PhysRevResearch.2.013133,bertulio}. Nevertheless, the importance of our work relies on what follows: first of all, the effective potential (\ref{effective}) is of pure quantum-mechanical origin since it is proportional to $\hbar$. It is due to the thin layer confinement of a quantum particle on a helicoid \cite{geometry,ferrari,paulicurved}. The term proportional to $l$ contribute as a quantum centrifugal potential, also proportional to $\hbar$. Both are due the combination of geometry and quantum dynamics. Secondly, such a particular geometry allows the engine to operate in a regime where the transverse direction $\rho$ across the helicoid remains constant ($r=1$). Consequently, the area of the sample may be kept unaltered. This actually correspond to keeping a constant volume of the sample, since it consists of a quasi-2DEG in reality. Notice that it is not possible to extract work in a classical Otto engine without volume modification of the sample since it works only for compressible fluids. The energy-level spacing of the system is the quantity that is being either compressed or stretched. For other values of $r$, the area of the helicoid and the energy level spacing are both modified.  The area of a finite helicoid is given by:
\begin{align}
	A&=\int_0^R d\rho\int_0^h dz\sqrt{1+\omega^2\rho^2}, \nonumber\\
	&= \frac{h}{2\omega}\left[\omega R\sqrt{1+R^2\omega^2}+\ln\left(\omega R+\sqrt{1+\omega^2R^2}\right)\right].\label{area}
\end{align}
As we change $\omega$, the height $h$ of the helicoid changes in order to keep the area constant. Taking $\omega_1 R=1$, it yields  $A_1=\frac{h_1}{2\omega_1}\left[\sqrt{2}+\ln\left(1+\sqrt{2}\right)\right]$. For $\omega_2 R=0.5$, we obtain $A_2=\frac{h_2}{2\omega_2}\left[\frac{1}{2}\sqrt{\frac{5}{4}}+\ln\left(\frac{1}{2}+\sqrt{\frac{5}{4}}\right)\right]$. Considering them as the areas at the end of the adiabatic strokes, we must have $h_1/h_2\approx1.1$. For a flat sample, only the convectional Otto machine is observed, since the only parameter to be modified is the transverse direction $\rho$ across a circle, for instance. We expect this also to happen for surface isometrics to it as a cylindrical bump or a surface with a catenary shape \cite{PhysRevB.80.153405}. On the other hand, a catenoid is locally isometric to the helicoid and it would provide similar geometry-induced quantum effects in heat machines. 

In Ref. \cite{jonas}, the authors addressed the quantum dynamics of one electron constrained to a curve or surface in order to obtain qualitative descriptions of the conduction electron's behavior in curved structures. Their studies can be useful to the investigations on quantum heat machines in distinct geometries.
\section{Concluding remarks}
In this contribution, we have addressed an Otto cycle heat machine driven by a 2DEG constrained to a twisted surface. We have employed the thin layer approach introduced by da Costa \cite{PRA1981}. By considering the adiabatic transformation of the number of complete twists per unit length of a helicoid, we showed that the machine performance can be altered, even improved. The realization of Otto machines with unaltered transverse direction $\rho$ across the helicoid is possible to be implemented in such a structure. The modifications in the wavefunctions of the electrons are due to the energy-level spacing of the system which is either compressed or stretched. This is not possible if one considers a flat sample(a circular one, for instance), unless in the case where the energies scale inhomogeneously by the introduction of a delta quantum potential \cite{PhysRevLett.120.170601}. Our results are obtained without the need for the application of external fields, which can add quantitative modifications to our observations. We expect our conceptual design of heat machines on a helicoid to motivate experiments probing quantum thermodynamics aspects of twisted materials. For instance, flexible semiconductive thin films \cite{flex} could offer a road to experimentally probe the features revealed here. 
If tables need to be extended over to a second page, the continuation
of the table should be preceded by a caption, e.g.~``{\it Table 2.}
$(${\it Continued}$)$''. Notes to tables are placed below the final
row of the table and should be flushleft.  Footnotes in tables should
be indicated by superscript lowercase letters and placed beneath the
table.

\section*{Acknowledgments}
We thank the Brazilian agencies CNPq, FAPEMA, and FAPEMIG for their financial support. Edilberto O. Silva acknowledges CNPq Grant 307203/2019-0, FAPEMA Grant UNIVERSAL-06395/22. C. Filgueiras and M. Rojas acknowledges FAPEMIG Grant APQ-02226-22. M. Rojas acknowledges CNPq Grant 317324/2021-7.  C. Romero and C. Filgueiras acknowledges CNPq grant 113754/2018-3.

\bibliographystyle{apsrev4-2}
%

\end{document}